\documentclass{article}

\PassOptionsToPackage{numbers, sort&compress}{natbib}
\usepackage[preprint]{neurips_data_2021}
\usepackage[utf8]{inputenc} % allow utf-8 input
\usepackage[T1]{fontenc}    % use 8-bit T1 fonts
\usepackage[colorlinks=true, linkcolor=blue, citecolor=green]{hyperref}    
\usepackage{url}            % simple URL typesetting
\usepackage{booktabs}       % professional-quality tables
\usepackage{amsfonts}       % blackboard math symbols
\usepackage{nicefrac}       % compact symbols for 1/2, etc.
\usepackage{microtype}      % microtypography
\usepackage{xcolor}         % colors

\usepackage{graphicx}       % Added by MK-Liu
\usepackage[english]{babel}
\title{Scalable Data Annotation Pipeline for High-Quality Large Speech Datasets Development} 

\author{%

  Mingkuan Liu\thanks{co-first authors, equal contribution.}\\ 
  Appen \\
  \texttt{mliu@appen.com} \\

  \And
  Chi Zhang\footnotemark[1] \\
  Appen \\
  \texttt{czhang@appen.com} \\

  \And
   Hua Xing \\
   Appen \\
  \texttt{hxing@appen.com} \\

   \AND
   Chao Feng \\
   Appen \\
  \texttt{cfeng@appen.com} \\

   \And
   Monchu Chen \\
   Appen \\
  \texttt{mochen@appen.com} \\

   \And
   Judith Bishop \\
   Appen \\
  \texttt{jbishop@appen.com} \\

   \And
   Grace Ngapo \\
   Appen \\
  \texttt{gngapo@appen.com} \\

}

\begin{document}

\maketitle

\begin{abstract} %

This paper introduces a human-in-the-loop (HITL) data annotation pipeline to generate high-quality, large-scale speech datasets. The pipeline combines human and machine advantages to more quickly, accurately, and cost-effectively annotate datasets with machine pre-labeling and fully manual auditing. Quality control mechanisms such as blind testing, behavior monitoring, and data validation have been adopted in the annotation pipeline to mitigate potential bias introduced by machine-generated labels. Our A/B testing and pilot results demonstrated the HITL pipeline can improve annotation speed and capacity by at least 80\% and quality is comparable to or higher than manual double pass annotation. We are leveraging this scalable pipeline to create and continuously grow ultra-high volume off-the-shelf (UHV-OTS) speech corpora for multiple languages, with the capability to expand to 10,000+ hours per language annually. Customized datasets can be produced from the UHV-OTS corpora using dynamic packaging. UHV-OTS is a long-term Appen project to support commercial and academic research data needs in speech processing. Appen will donate a number of free speech datasets from the UHV-OTS each year to support academic and open source community research under the CC-BY-SA license. We are also releasing the code of the data pre-processing and pre-tagging pipeline under the Apache 2.0 license to allow reproduction of the results reported in the paper. \protect\footnote[1]{Code and data are available in \url{https://github.com/Appen/UHV-OTS-Speech}}

\end{abstract} 

\section{Introduction}
\label{introduction}

Rapid advances in deep learning technology over the past decade have led to frequent improvements in state-of-the-art results for speech tasks on standard benchmarks. Yet the performance of speech applications deployed in the real world lags well behind the published benchmark results, especially when the real-world use cases are complicated by noise, accents, different domains, etc.

There are two possible reasons for this performance gap. First, the datasets used for benchmark testing may not sufficiently represent complex real-world scenarios. Second, the datasets used to train the AI models deployed in speech applications may lack sufficient coverage of these real-world cases.

Standard mainstream speech recognition benchmark corpora have not changed much in decades. The Wall Street journal corpus \cite{wsjcorpus} , Switchboard corpus \cite{swithboardcorpus} , and Fisher corpus \cite{David04thefisher} are around 20 years old. The Librispeech corpus \cite{librispeech} is newer and widely used but is based on audio books. The speaking styles in these corpora tend heavily to narrative or scripted speech, and the accents and levels of background noise are not reflective of the real-world scenarios in which ASR is deployed, which may involve face-to-face conversation, spontaneous speaking styles, diverse accents and noisy environments.

It is expensive and time-consuming to develop large-scale, high-quality speech datasets in the traditional way, by manually collecting and annotating audio data with multiple passes. The prohibitive investment cost and slow delivery time have contributed significantly to advanced speech corpus development lagging far behind advances in speech algorithms.

This paper introduces a human-in-the-loop data annotation pipeline to generate high-quality, large-scale speech datasets more quickly, accurately, and cost effectively. The pipeline combines human and machine advantages with machine pre-labeling, intelligent quality control, and fully manual auditing. 

The contributions of our work are as summarized below:

\begin{itemize}
    \item We have created a human-in-the-loop data annotation pipeline that combines human and machine advantages to annotate datasets more quickly, accurately and cost effectively, using machine pre-labeling and fully manual auditing. We expect this development to be of interest to industry data consumers.
    \item We have developed multiple intelligent quality control mechanisms such as blind testing, behavior monitoring, and data validation to ensure annotation quality and mitigate potential bias introduced by machine-generated labels.
    \item We are generating continuously growing, ultra-high volume off-the-shelf (UHV-OTS) speech corpora for multiple languages, and we are offering these together with dynamic packaging to facilitate selection of customized subsets of the corpora. The UHV-OTS contains rich speaker variation and meta-information about speaker gender, and accent, as well as domain, topic, and background noise at the utterance, session, speaker, and dataset levels.
    \item Additionally, this project could support crowd workers with a stable income stream, and we propose to provide academic and open source research communities with regularly donated speech datasets, free of charge, under a CC-BY-SA license.
\end{itemize}

\section{Related work}
\label{relatedwork}

Academia, industry and the open source community have recently created a range of speech corpora for model training and benchmark testing, especially for English language speech recognition. 

\begin{itemize}
    \item TED-LIUM corpus \cite{conf/lrec/RousseauDE12} is based on narrated TED talks. It has less than 1000 audio hours. 
    \item Librispeech \cite{librispeech} has 1000 hours of data collected from audio books, with a single speaker reading a text in a quiet environment.
    \item Mozilla Common Voice \cite{commonvoice} is a 10000 hr public domain corpus covering multiple languages with a narrative speaking style. 
    \item SPGIspeech \cite{spgispeech} has 5000 hours of audio transcribed from earnings calls. This business domain dataset contains both spontaneous and narrative speaking styles.
    \item People's Speech corpus \cite{peoplespeech} contains 31400 hours of mostly English audio data scraped from the internet. This corpus has a CC-BY-SA license. It uses forced alignment of the audio against transcripts to create the training dataset. Lacking a human audit of the transcriptions, the overall labelling quality is not high. However, it is an evolving corpus and includes multiple domains and speaking styles. 
    \item Gigaspeech \cite{gigaspeech} has 10000 hours of English audio data. Similar to the People's Speech corpus, forced alignment is used to create the speech dataset. It is also an evolving speech recognition corpus including multiple domains and speaking styles. 
\end{itemize}

While some researchers are building better-labeled training datasets for supervised speech recognition models, other researchers are exploring options that require a reduced quantity of transcribed audio data. \cite{fbhubert}, \cite{fbselfsupervised}, and \cite{fbwav2vec} use unsupervised and self-supervised learning approaches to train speech recognition systems with mostly raw, unlabeled audio and very little transcribed audio data.   

Comparing with related works, the UHV-OTS corpora have the advantages of: 1) high quality transcriptions as a result of a fully manual audit, 2) rich and detailed metadata to support the training of models not only for speech recognition, but also speaker diarization, speaker identification, accent and gender detection, 3) highly varied speakers, domains and speaking styles to ensure dataset is representative of complex, real-world environments, and 4) a streamlined and optimized pipeline that can be easily scaled up to build high volume datasets across many languages with significantly reduced development costs.  

The work presented in this paper complements the speech corpora cited above. Raw audio data from the People's Speech dataset \cite{peoplespeech} could potentially be included in the UHV-OTS corpora's audio dataset, with potential to improve the label quality of the People's Speech dataset through the HITL data annotation pipeline. The diversity of speakers and audio conditions in the UHV-OTS can benefit self-supervised approaches such as \cite{fbhubert}, \cite{fbselfsupervised}, \cite{fbwav2vec}. By introducing more diversity, our data can help self-supervised ASR models be more robust in complex real-world scenarios.

\section{Speech corpora design}
\label{Corpora Design}

A speech training dataset combines speech audio, corresponding transcriptions, acoustic event tagging and speech/audio/speaker-related metadata such as accent, background noise, topic, domain, gender etc. With these detailed and precise annotations on the audio data, data scientists can use the speech datasets to develop machine learning applications such as speech recognition, speech synthesis, speech/speaker segmentation and diarization, speaker identification, language/accent identification, gender classification, etc.

Our UHV-OTS speech corpora provide detailed metadata at four different levels. 

\begin{itemize}
    \item Speech dataset level metadata - this metadata contains language, accent, speaker-related demographic information, topics and audio distribution statistics at the level of the entire dataset.

    \item Session level metadata - an audio session is a group of related utterances, for example, an hour-long audio clip of an interview. All session level metadata contains information such as session\_id, speakers, audio\_path, duration, utterance\_ids\_list, domains, topics, and accents.

    \item Utterance level metadata - an utterance is a short audio clip, usually comprising a single spoken sentence of no more than 20 seconds' duration. All metadata at the utterance level has information such as path to the audio file, speaker ID, speaker accent, speaker gender, text transcription, audio length, session\_id, topics and background noise type.

    \item Speaker level metadata - every speaker has a metadata file which contains information about speaker ID, accent, gender, language, list of utterance\_ids, list of session\_ids, and the duration of audio from this speaker. To ensure highly diversified speakers in the corpora, audio from the same speaker is limited to less than 60 minutes. 

\end{itemize}

Appendix \ref{dataset format} dataset format section contains detailed samples of the JSON files for each level of metadata.

\subsection{Dynamic packaging}

Dynamic packaging enables the streamlined production of customized speech datasets, which are a smaller subset of the corpora. Each customized dataset comprises selected distributions of speaker gender, accent, topic/domain, noise level, etc. from the very large pool of UHV-OTS corpora. The dynamic packaging feature allows data consumers to obtain maximum value for their budget for specific use cases.  

All speech datasets that are dynamically packaged out of the UHV-OTS corpora will follow the data format described in \ref{dataset format} section. 

Naming of the speech datasets for commercial usage will follow the pattern: UHV-OTS-Commercial-\{locale\}-\{domain\}-\{projectname\}-\{releasedate\}.

Naming of regularly donated speech datasets will follow the pattern: UHV-OTS-Research-\{locale\}-\{domain\}-\{releasedate\}.

\section{Speech dataset development pipeline}
\label{pipeline}

The traditional method of developing high quality speech corpora is expensive and time-consuming due to its reliance on several passes of manual data collection, pre-processing, annotation, and post-processing.

Earlier experimental results in \cite{asrcmu} indicated that transcribers starting with ASR pre-labeling performed worse than those starting from scratch unless the ASR system is sufficiently accurate (Word Error Rate under 30\%). Our internal A/B test results, confirmed through multiple testing rounds, also independently verified those observations. We found that when machine pre-labeling accuracy was less than 70\%, machine pre-labeling can negatively impact human annotator speed and accuracy. However, when machine pre-labelling accuracy was greater than 85\%, machine pre-labeling can significantly improve human annotator speed and accuracy.

Based on the above findings, we've included several, high accuracy, machine pre-labeling components into our human-in-the-loop speech data annotation pipeline to generate high-quality, large-scale speech coropra more quickly, accurately, and cost effectively. This pipeline combines human and machine advantages with machine pre-labeling, intelligent quality control, and fully manual auditing. Our A/B test and pilot running have demonstrated its effectiveness: annotators' efficiency and capacity can be improved by at least 80\% and the quality is comparable to or higher than manual double pass annotation. 

\begin{figure}[h]
  \centering
  %\fbox{\rule[-.5cm]{0cm}{4cm} \rule[-.5cm]{4cm}{0cm}}
  \includegraphics[width=13cm]{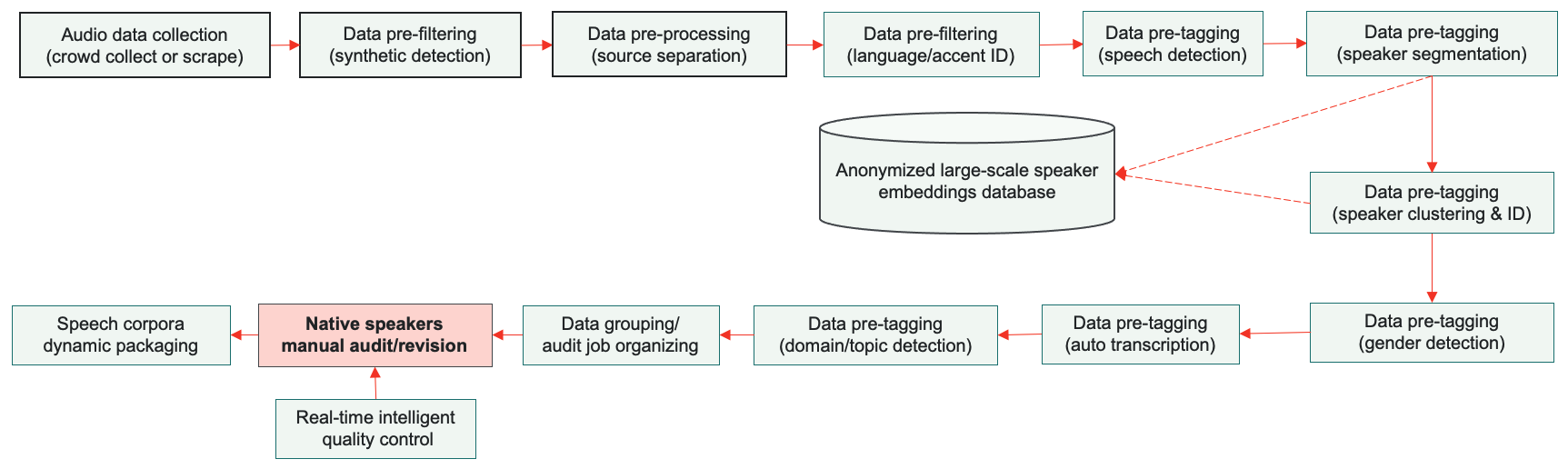}
  \caption{HITL speech datasets development pipeline}
\end{figure} 

\subsection{Audio data collection: crowd collection and web scraping}

The raw audio data sources for the UHV-OTS corpora can come from crowd collection or web scraping. Crowd collection is slow and expensive, it's used when web scraping method can't cover the needed use cases.

There is an almost unlimited amount of raw audio or video data available on the internet, generated by millions of speakers. The web scraped raw audio data used in our UHV-OTS corpora is from the internet sources with license types that allow commercial use of the data. In our data annotation pipeline, original taggings or transcriptions are not required, which increases the volume of audio resources available for use in UHV-OTS datasets compared with others like \cite{peoplespeech} and \cite{gigaspeech}. UHV-OTS audio files are stored in 16000 Hz 16bit linear PCM wav format and mp3 format. 

\subsection{Data pre-processing: source separation}
Audio scraped from the internet often contains music segments or background music in speech segments. Music segments can be excluded by using the speech detection algorithm described in section \ref{pre-tagging:speech detection}. However, speech with background music will not always be detected and will negatively affect the performance of further processing steps. In \cite{spleeter2020}, the spleeter source separation tool was proposed and proven to be effective in separating vocal and musical signals. We adopted the spleeter tool in our pipeline to pre-process the downloaded audio data. The resulting separated vocal signal is used for further processing, tagging and transcription to ensure higher accuracy.

\subsection{Data pre-filtering: synthetic speech detection}
With recent advances in synthetic speech technology, there are substantial amounts of internet audio data generated by machines instead of humans. To ensure our corpora contains high quality speech data, the synthetic speech audio needs to be filtered out from the web scraped raw data. 

In \cite{Hua_2021}, a light-weight end-to-end neural network was proposed that achieved the state-of-the-art synthetic speech detection result on the ASVspoof2019 \cite{todisco2019asvspoof} Logical Access dataset. The "Automatic Speaker Verification Spoofing and Countermeasures" (ASVspoof) challenges are bi-annual research challenges to accelerate anti-spoofing research. The Logical Access partition of the ASVspoof2019 dataset contains the synthesized/converted speech to spoof a speaker. The work in \cite{Hua_2021} achieved synthetic speech detection EER as low as 2.16\% on in-domain testing data and 1.95\% on cross-domain data. This algorithm was implemented in our data annotation pipeline. A model trained with the same datasets as in \cite{Hua_2021} was used to detect extracted audio containing spoofed speech, which was excluded from further processing.

\subsection{Data pre-filtering: language \& accent identification} \label{pre-filtering:lang&acct}

\begin{figure}[h] 
  \centering
  \includegraphics[width=10cm]{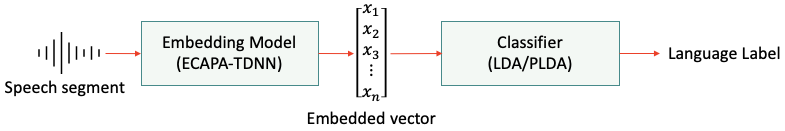}
  \caption{Flow diagram of ECAPA-TDNN + LDA/PLDA language ID algorithm}
  \label{fig:emb_clas}
\end{figure} 

Accent and language are important metadata for speech corpora. We apply language and accent identification to pre-filter the raw audio data and ensure that the data is correctly routed to the corresponding language/accent data processing pipeline.

A language ID algorithm of embedding-plus-classification framework was implemented for our annotation pipeline, as illustrated in Fig. \ref{fig:emb_clas}. The ECAPA-TDNN embedding method \cite{desplanques2020ecapa} was adopted in our pipeline implementation, which is a variation of the conventional x-vector \cite{snyder2018x} embedding model. The ECAPA-TDNN model was trained with the VoxLingua107 dataset \cite{valk2021slt}, and the language ID algorithm achieved 93\% accuracy on the VoxLingua107 dev set. In the future, as increasing quantities of labeled data become available in our UHV-OTS corpora, we will continuously optimise the model to further improve performance. 

Accent identification is more challenging than language identification due to lack of available datasets and difficult to distinguish subtleties between some accents. We've adopted the x-vector plus LDA/PLDA framework to detect twenty-two different English accents using proprietary data. Our current accent detection accuracy is 75\%. Similar to the language ID algorithm, we will continuously optimise the accent ID model to improve performance as increasing quantities of labeled accent data become available in our UHV-OTS corpora.

\subsection{Data pre-tagging: speech/non-speech audio segmentation}
\label{pre-tagging:speech detection}
Raw audio data may contain non-speech signals such as silence, background noise, music etc. It is important to tag this information when developing speech corpora for training audio-related machine learning models.  The \textit{inaSpeechSegmenter} \cite{ddoukhanicassp2018} speech detection module is adopted in our data annotation pipeline to segment the input audio into homogeneous zones of speech, music, and noise.

The \textit{inaSpeechSegmenter} system won the first place in the Music and/or Speech Detection in Music Information Retrieval Evaluation eXchange 2018 (MIREX 2018) \cite{2018MIREX}. This module also achieved 97.5\% detection accuracy with an average boundary mismatch of 97ms at Appen's proprietary testset. 

\subsection{Data pre-tagging: speaker segmentation}

Multi-speaker conversational speech is very common in real-world application scenarios such as conference meetings, call center conversations, broadcast shows etc. It's important to partition the input audio stream into homogeneous segments according to speaker identity. This process is called speaker segmentation, which answers the question "who spoke when". 

\begin{figure}[h] 
  \centering
  \includegraphics[width=10cm]{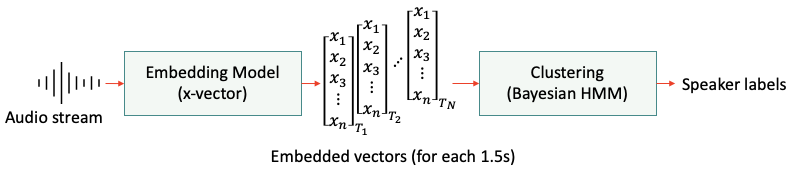}
  \caption{Flow diagram of speaker segmentation algorithm}
  \label{fig:spkr_seg}
\end{figure}

The BUT speaker diarization framework \cite{landini2021analysis} is adopted in our data annotation pipeline for speaker segmentation and speaker clustering purposes. The speaker diarization framework generally involves an embedding stage followed by a clustering stage, which is illustrated in Fig. \ref{fig:spkr_seg}. 

We tested the pipeline with VoxConverse corpus \cite{chung2020spot}, which is an audio-visual diarization dataset consisting of over 50 hours of multi-speaker clips of human speech, extracted from videos collected on the internet. The DER achieved on VoxConverse using the BUT system is 4.41\%, which is consistent with the result in \cite{landini2021analysis}.

\subsection{Data pre-tagging: speaker clustering \& identification}\label{pre-tagging:speakderid}

To ensure rich diversity of speakers in our UHV-OTS corpora, we limited the maximum total speech duration from any particular speaker to 60 minutes. To achieve this, however, we need to identify the same speaker across multiple audio sessions. Speaker clustering, speaker identification, and an on-going, anonymized speaker database are used to enforce this limit. The diagram below shows our workflow. 

\begin{figure}[h]
  \centering
  \includegraphics[width=13cm]{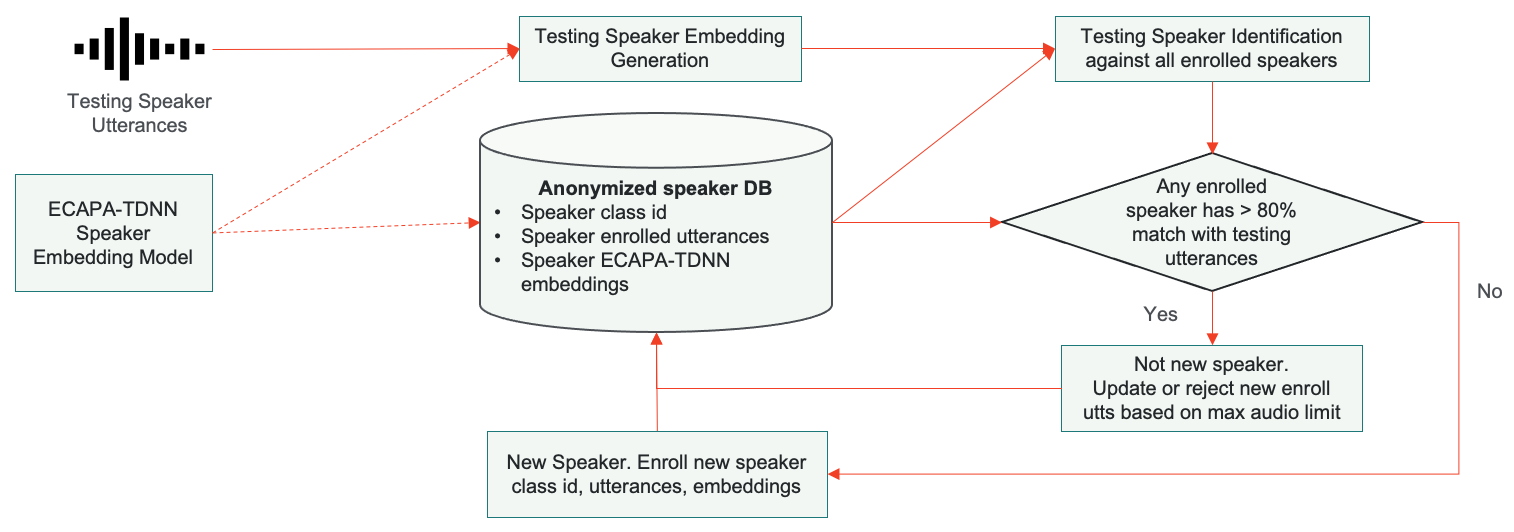}
  \caption{Speaker clustering, identification and anonymized speakers DB enrollment workflow}
\end{figure} 

The ECAPA-TDNN embedding algorithm \cite{desplanques2020ecapa} is adopted to generate speaker embeddings, which is used for speaker identification purpose. The output of the previous speaker segmentation step is a set of audio segments, where each set belongs to a single speaker from a single speech session. Next, we aim to detect whether each speaker already exists in the anonymized speakers DB (i.e., if the testing speaker matches one of our enrolled speakers). If we find the test segments are produced by an enrolled speaker, the enrolled set will be updated by adding these segments to the corresponding class. Otherwise, the new segments will form a new speaker id class and these audio segments will be added to the enrolled data associated with the new speaker id. A pre-trained embedding model by SpeechBrain toolkit is adopted in our pipeline, which produces EER of 0.7\% on VoxCeleb 1 dataset \cite{nagrani2017voxceleb}.

\subsection{Data pre-tagging: speaker gender detection}
\label{pre-tagging:gender}
Speaker gender information is required to maintain a balanced gender representation in the speech corpora. An  x-vector embedding model plus Multi-layer Perceptron (MLP) classifier framework is implemented in our annotation pipeline for speaker gender detection. Our gender detection model achieved 99.85\% accuracy on VoxCeleb1 testing set \cite{nagrani2017voxceleb}. 

\subsection{Data pre-tagging: transcribe with ASR}\label{pretagging:asrtranscribe}

The automatic speech recognition (ASR) module is used in our annotation pipeline to pre-label hypothesized transcriptions for raw audio data. Those pre-labeled transcriptions will be audited and further corrected by downstream, native human annotators. Kaldi toolkit \cite{Povey11thekaldi} and the Chain model recipe \cite{Povey+2016} were adopted as our ASR module in the annotation pipeline. Multiple speech dataset sources covering multiple domains were used to train our ASR model, similar to the work in \cite{DBLP:journals/corr/abs-2010-11745}  \cite{DBLP:journals/corr/abs-2104-02133}, but with a different model infrastructure. 

Multiple rounds of internal A/B testing demonstrated that ASR-assisted transcription of speech, with an ASR WER of 10.8\%, can improve productivity by 112\% when compared with transcribing from scratch. This result is in line with \cite{asrcmu} earlier observation.

\subsection{Data pre-tagging: domain and topic detection} \label{pre-tagging:topic}

Domain and topic are important meta-information for speech corpora. Based on our market needs analysis, we have defined a set of 20 top topics and categories, namely Clothing, Culture, Education, Finance, Food, Health ,History, Hospitality, Information and Technology, Insurance, Legal, Leisure time, Entertainment(TV, Film, Music, News), Retail, Social networks, Sports, Telecommunication, Travel/Holiday, Weather, Work.

We use the above categories as seed keywords to scrape relevant video and audio files from the internet. If feasible and allowable, any metadata associated with the video/audio is also downloaded, and where available, topic and category tags extracted from this metadata are included in final delivery. When the topic and category tags are not available, a topic classification module will be adopted in the pipeline to estimate topics based on the full audio session's ASR hypothesis generated in previous step.

\subsection{Automated labeling module performance summary} 
To better illustrate the quality of tags from automated labeling modules used in the annotation pipeline, we summarized the algorithms used and their performance in the table below.

\begin{table}[h]
    \centering
    \scalebox{0.7}{
    \begin{tabular}{|c|c|c|c|c|}
    \hline
         \textbf{Module} & \textbf{Algorithm} & \textbf{Train Set} & \textbf{Test Set} & \textbf{Metric} \\\hline\hline
         Synthetic detection & Res-TSSDNet \cite{Hua_2021} & ASVspoof2019 \cite{todisco2019asvspoof} & ASVspoof2015 & EER: 1.95\% \cite{Hua_2021} \\\hline
         Language ID &  ECAPA-TDNN + LDA/PLDA \cite{desplanques2020ecapa}& VoxLingua107 \cite{valk2021slt} & VoxLingua107 dev & Accuracy: 93\% \\\hline
         22 English accent ID &  ECAPA-TDNN + LDA/PLDA \cite{desplanques2020ecapa}& Proprietary dataset & Proprietary dataset & Accuracy: 75\% \\\hline
         Speech detection & CNN + HMM \cite{2018MIREX} & GTZAN \cite{1021072} etc. & In-house set & Accuracy: 97.5\% \\\hline
         Speaker segmentation & BUT system \cite{landini2021analysis} & VoxCeleb2 \cite{nagrani2017voxceleb} & VoxConverse & DER:4.41\% \\\hline
         Speaker ID &  ECAPA-TDNN +LDA/PLDA \cite{desplanques2020ecapa} & VoxCeleb1 & VoxCeleb1 test & EER:0.7\% \\\hline
         Gender detection & X-vector + MLP & VoxCeleb1,2 & VoxCeleb1 test & Accuracy: 99.85\% \\\hline
         ASR & Chain model \cite{Povey+2016} & 11 corpora & Librispeech test-clean \cite{Panayotov2015LibrispeechAA} & WER: 2.8\% \\\hline
    \end{tabular}
    }
    \caption{Automatic modules algorithms and performance summary}
    \label{tab:my_label}
\end{table}
 
\section{Quality control mechanisms to ensure accurate annotation}

High quality labeled datasets are critical for developing and evaluating supervised machine learning models. All annotations in our UHV-OTS speech corpora are fully manually audited and revised by native speakers to ensure the highest quality.

Human auditing of machine pre-labeled data can improve annotation speed. However, researchers \cite{asrmsft} have expressed a valid concern that machine-generated labels could introduce bias. They argue that human annotators primed with generally reliable machine-generated labels get habituated to seeing correct labels and therefore trust the pre-labels more than they deserve. 

To mitigate such concerns, \cite{apnpatent1} have proposed multiple quality control mechanisms which have been implemented in our annotation pipelines. Internal A/B testing and pilot results have demonstrated the effectiveness of these quality assurance methods, with quality being comparable to or higher than manual double pass annotation.

The major components of our quality control process are shown in Figure 5 below. 

\begin{figure}[h]
  \label{QCdiagram}
  \centering
  \includegraphics[width=13cm]{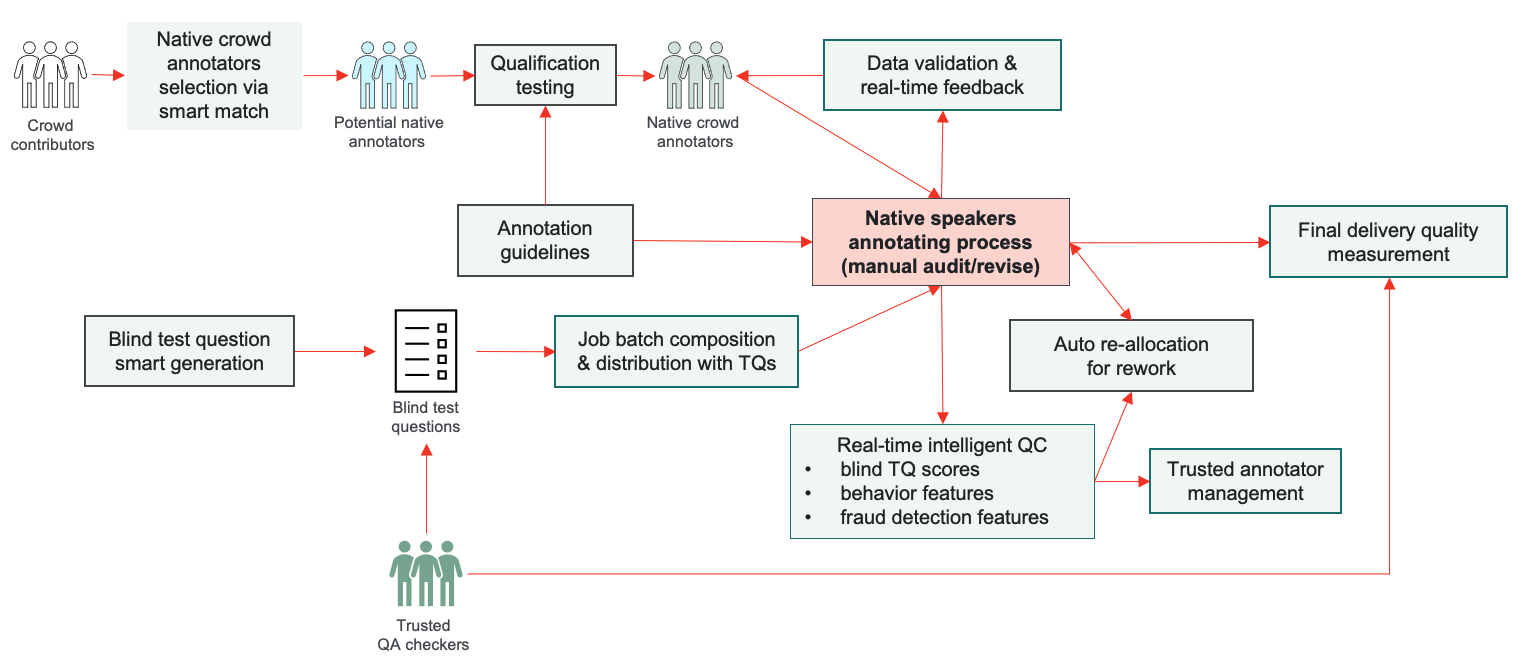}
  \caption{Intelligent quality control mechanisms in the annotation pipeline}
\end{figure} 

\subsection{Annotator selection \& guideline training}

Typically, our data annotation process is completed online by crowd annotators using Appen's proprietary tools. We initially select annotators that are native speakers, who are then required to complete qualification tests in their chosen language and dialect or accent. They are tested on their spelling accuracy, their ability to complete relevant tasks while adhering to guidelines, and their knowledge of terms specific to the task locale. Clear and detailed annotation guidelines ensure annotation correctness and consistency across the data set. Only high-scoring, native speaking annotators are invited to join annotation jobs.

\subsection{Blind testing based quality control}

In a blind experiment, information which may influence experiment participants is withheld until after the experiment is complete. Effective blinding can reduce or eliminate experimental bias arising from participants' expectations, observers' effects on participants, observer bias, confirmation bias, and other sources of experimental bias. 

In our annotation pipeline, we first generated a set of blind test questions (TQ) with ground truth answers verified by trusted quality checkers. \cite{apnpatent2} describes an efficient mechanism to generate high quality blind test questions quickly in a scalable manner at low cost. For every annotation assignment (i.e., a certain number of units to be annotated), a few blind test questions are randomly injected into the assignment, such that the crowd annotator can't tell which unit is a blind test question or a regular work item. When the annotator submits the assignment result, the blind TQ-based quality score for this particular annotator is automatically computed and updated by comparing the answer submitted against the ground truth. If a crowd annotator's TQ quality score falls below a certain threshold, they will be removed from the job. In addition, all units annotated by this annotator will be automatically recycled and re-distributed to higher quality annotators.

\subsection{Behavior monitoring based quality control}

A behavioral approach to quality control looks at annotators' behavioral traits to determine the quality of the judgments without validating against the ground truth. This is an additional mechanism to further improve annotation quality and mitigate the potential bias introduced by machine-generated labels. \cite{internqc1} demonstrated a high correlation between behavioral features and annotation quality.

In the data annotation pipeline, multiple annotator behavioral signals were monitored to ensure annotation quality. The well-known priming effect occurs when a response to the second stimulus is affected by the first stimulus. When presenting machine predictions in a machine-assisted transcription job, an annotator may tend to “confirm” what they read (first stimulus) in relation to what they hear in the audio clip (second stimulus). Behavioral features such as the number of edits, time spent on editing and time spent listening are reliable proxies to determine their transcription performance. Furthermore, a hard check on their listening behaviors (e.g. whether the annotator finishes listening to the whole clip or not) has been an easy and very effective mechanism to flag low quality annotators.   

\subsection{Real-time data validation \& feedback}

To further help annotators produce high quality annotations, real-time data validation and feedback mechanisms with flexible configuration are also implemented in the annotation pipeline. These mechanisms validate annotator submissions against annotation guidelines, including spelling and formatting. A validation error message is sent back to the annotator in real-time to help prevent future errors of that kind. 

\subsection{Final delivery quality measurement}

The ISO-9001 quality management requirements \cite{tmp} are followed in our data annotation pipeline. Before final packaging and delivery of annotation results, a statistically significant number of annotated units is randomly sampled and audited by trusted quality checkers. The quality of the final delivery is assessed against the threshold required to ensure it meets the desired accuracy. If annotation quality falls below the threshold in the final quality assessment, a statistically determined proportion of units is reworked and re-audited until the dataset as a whole exceeds the accuracy threshold.
 
\section{Discussion }

The UHV-OTS speech corpora development is an ongoing, long-term Appen project to support commercial and academic research data needs for tasks related to speech processing.  

The goal of the UHV-OTS project is to develop speech corpora with good coverage of complex real-world cases, so that the AI applications trained with these corpora will perform better in challenging, real-world deployments. Based on our market needs analysis, 20 popular vertical domains including Culture, Education, Finance, Food, Health, Insurance, Legal, Entertainment, Retail and Sports will be covered initially by the UHV-OTS corpora. In the UHV-OTS project road map, Appen is establishing a delivery capacity of 10000+ hours per year of speech datasets for each language. Target languages to cover include English, French, Italian, German, Spanish, Arabic, Chinese, Japanese and Korean. However, the roadmap will be regularly adjusted based on industry and academic demand.

The continuous nature of UHV-OTS corpora development will allow us to serve commercial and academic needs more quickly and effectively. Dataset consumers can visit \url{https://appen.com/off-the-shelf-datasets/} and inform us of their specific dataset needs. Appen will consolidate these inputs and adjust our UHV-OTS delivery pipeline accordingly to ensure our speech corpora meets the most common needs of consumers. We also expect that the longevity of this project will help crowd annotators to maintain a stable income from the annotation work, providing an opportunity for greater income security.  

The dynamic packaging feature of UHV-OTS will allow data consumers to optimize their budget for specific use cases. Each custom dataset will comprise of consumer-requested distributions of audio hours, speaker genders, accents, topics and domains, noise levels, etc., from the very large pool of UHV-OTS corpora. 

Finally, Appen will donate a certain number of speech datasets each year from the UHV-OTS corpora to support academic and open source community research. These free datasets will be published under the CC-BY-SA license and will be downloadable from Appen's \url{https://appen.com/open-source-datasets/} website.

\newpage

\bibliography{references}

\appendix

\section{Appendices}

\subsection{Dataset format with detailed samples}
\label{dataset format}

The UHV-OTS speech corpora combine speech audio, corresponding transcriptions, acoustic event tagging and speech/audio/speaker-related metadata such as accent, background noise, topic, domain, gender etc. Those detailed metadata are presented at utterance, session, speaker, and dataset levels as below sample in the format of JSON. 

\subsubsection{Dataset level meta data}
A typical sample UHV-OTS speech dataset level metadata JSON file as below
\begin{verbatim}
UHV-OTS-Commercial-enus-general-lighthouse-2021XXXX.json=
{
   "speechdb_name":"UHV-OTS-Commercial-enus-general-lighthouse-2021XXXX",
   "language":  "english",
   "accent": "en-us",
   "duration_in_hours": 210,
   "speakers_cnt":  310, 
   "utterances_cnt": 53418,
   "Topics_by_hours": {               
        "sports":20.5, 
        "culture":53.5, 
        "education":46.4, 
        "finance":30.6, 
        "food":59
    },
   "Topics_by_speakers": {               
        "sports":52, 
        "culture":76, 
        "education":50, 
        "finance":48, 
        "food":84
    },
   "gender_dist_by_hours": { "male": 117.5, "female": 92.5}, 
   "gender_dist_by_speakers": { "male": 164, "female": 146}, 
   "noisetype_dist_by_hours":  {"clean": 35, "noisy": 105, "music":70}, 
   "sampling_rate": 16000, 
   "sampling_bit": 16, 
   "audio_channels": 1 
} 
\end{verbatim}
\newpage

\subsubsection{Session level meta data}
A typical sample session level metadata JSON file with session\_id="asd123efs" as below.
\begin{verbatim}
asd123efs.json=
{ 
    "session_id": "asd123efs",
    "audio_path": "/audio-session/asd123efs.mp3", 
    "duration_in_minutes": 35.7, 
    "utterance_ids_list": ["asd123efs-1", ..., "asd123efs-28" ], 
    "speakers" : ["sddseewsf32sxeor", "sadflk23laevs"], 
    "session brief title": "nba sports news westbrook", 
    "domains": ["sports"],  
    "topics": ["sports", "basketball", "nba"], 
    "language": "English", 
    "accent": "en-us", 
    "noise_background": "noisy", 
    "sampling_rate": 16000, 
    "sampling_bit": 16
} 
\end{verbatim}

\subsubsection{Utterance level meta data}
A typical sample utterance level metadata JSON file with utterce\_id="asd123efs-123" as below.
\begin{verbatim}
asd123efs-123.json=
{ 
    "utterce_id": "asd123efs-123", 
    "speaker_id": "sddseewsf32sxeor", 
    "session_id": "asd123efs", 
    "audio_path": "/audio-utterance/asd123efs/asd123efs-123.mp3",
    "duration_in_seconds": 15.3,
    "domains": ["sports"],
    "topics": ["sports", "basketball", "nba"],
    "transcription": "Westbrook had thirty five points, fourteen 
      rebounds and twenty one assists to lead Washington to a win."
    "language": "English",
    "accent": "en-us",
    "gender":  "male",
    "noise_background": "noisy", 
    "sampling_rate": 16000,
    "sampling_bit": 16
}     
\end{verbatim}

\subsubsection{Speaker level meta data}
A typical sample utterance level metadata JSON file with speaker\_id="sddseewsf32sxeor" as below.
\begin{verbatim}
sddseewsf32sxeor.json=
{ 
	"speaker_id": "sddseewsf32sxeor", 
	"utterce_ids_list": ["asd123efs-11", "asd123efs-23","weadsffdsa-321",... ], 
	"context_ids_list": ["asd123efs", "weadsffdsa"], 
	"duration_in_minutes": 45.3,  
	"language": "English", 
	"accent": "en-us", 
	"gender":  "male", 
}
\end{verbatim}

\end{document}